\documentclass[journal]{IEEEtran}
\usepackage{xcolor}
\usepackage{graphicx}
\usepackage{booktabs}
\usepackage{multirow}
\usepackage{tabularx}
\usepackage{array}
\usepackage{threeparttable}
\usepackage{dblfloatfix}
\usepackage{tikz}
\usepackage{lipsum}
\usepackage{makecell}
\usepackage[caption=false,font=footnotesize]{subfig}
\usepackage[most]{tcolorbox}
\usepackage{pifont}
\usepackage{adjustbox}
\usepackage{ragged2e}
\usepackage{cite}

\DeclareRobustCommand{\circledsubfiglabel}{%
    \ifcase\value{subfigure}%
    \or \ding{172}
    \or \ding{173}
    \or \ding{174}
    \or \ding{175}
    \else \arabic{subfigure}%
    \fi
}

\newlength{\rowdescw}
\newlength{\rowgapw}
\newlength{\rowcontentw}
\newlength{\figH}

\newcommand{\rowdesctext}[2]{%
    \centering\scriptsize\textbf{#1}\par\vspace{0.2em}#2%
}

\newtcolorbox{subfigbox}{
    enhanced,
    width=\linewidth,
    colback=white,
    colframe=black!65,
    boxrule=0.55pt,
    arc=1.5mm,
    left=1.0mm,
    right=1.0mm,
    top=1.0mm,
    bottom=1.0mm,
    before skip=0pt,
    after skip=0pt
}

\newtcolorbox{refbox}{
    enhanced,
    width=\linewidth,
    colback=white,
    colframe=red!65,
    boxrule=0.55pt,
    arc=1.0mm,
    left=1.0mm,
    right=1.0mm,
    top=0.8mm,
    bottom=0.8mm,
    before skip=0pt,
    after skip=0pt,
    fontupper=\footnotesize
}

\newtcolorbox{limbox}{
    enhanced,
    width=\linewidth,
    colback=red!3,
    colframe=red!65,
    boxrule=0.55pt,
    arc=1.0mm,
    left=1.0mm,
    right=1.0mm,
    top=0.8mm,
    bottom=0.8mm,
    before skip=0pt,
    after skip=0pt,
    fontupper=\footnotesize
}

\newcommand{\dashsep}{%
    \par\vspace{0.18em}
    \noindent
    \tikz{\draw[gray!45,dashed,line width=0.35pt] (0,0) -- (\linewidth,0);}
    \par\vspace{0.18em}
}

\newcommand{\rowdashsep}{%
    \par\vspace{0.55em}
    \noindent
    \tikz{\draw[gray!55,dashed,line width=0.45pt] (0,0) -- (\textwidth,0);}
    \par\vspace{0.15em}
}

\newcommand{\fixedfig}[1]{%
    \centering
    \includegraphics[
        width=\linewidth,
        height=\figH,
        keepaspectratio
    ]{#1}
}

\newcommand{\workline}[1]{%
    {\scriptsize #1}
}

\renewcommand{\rowdesctext}[2]{%
    \centering\scriptsize\textbf{#1}\par\vspace{0.2em}#2%
}

\newcommand{\paradigmpanel}[5]{%
\subfloat[#1]{%
\begin{minipage}[t]{0.485\linewidth}
\begin{subfigbox}
    \fixedfig{#2}

    \dashsep

    \begin{refbox}
        \centering
        {\scriptsize\textbf{\textcolor{blue!45!black}{Representative works:}}}\\[-0.25em]
        \workline{#3}
    \end{refbox}

    \dashsep

    \begin{limbox}
        {\scriptsize\textbf{\textcolor{red!75!black}{Limitations:}}}\\[-0.25em]
        {\scriptsize\RaggedRight #4}
    \end{limbox}
\end{subfigbox}
\label{#5}
\end{minipage}%
}%
}

\DeclareRobustCommand{\figthreecircledsubfiglabel}{%
    \ifcase\value{subfigure}%
    \or \ding{172}
    \or \ding{173}
    \or \ding{174}
    \or \ding{175}
    \else \arabic{subfigure}%
    \fi
}

\title{Rethinking Edge-Device Collaborative LLM Inference for 6G AI Edge: An Empirical Evaluation}
\author{Cheng Wu, Xinchen Lyu, Chenshan Ren, Ziyao Yan, Qimei Cui, Xiaofeng Tao, and Yang Yang, \IEEEmembership{Fellow,~IEEE}%
\thanks{C. Wu, X. Lyu, Z. Yan, Q. Cui, and X. Tao are with the National Engineering Research
Center for Mobile Network Technologies, Beijing University of Posts and
Telecommunications, Beijing 100876, China.}%
\thanks{C. Ren is with the Key Laboratory of Ethnic Language Intelligent Analysis
and Security Governance of MOE, Minzu University of China, Beijing
100081, China.}%
\thanks{Y. Yang is with Shanghai Center of Hong Kong University of Science and Technology, Shanghai 200232, China.}
\
}
\begin{document}

\maketitle

\bstctlcite{IEEEexample:BSTcontrol}

\begin{abstract}
6G ubiquitous intelligence requires generative AI capabilities that far exceed the compute, memory, and energy budgets of mobile terminals.
By pushing large language model (LLM) inference to the proximate network edge, 6G is shifting to the AI Edge era via four edge-device collaborative paradigms: edge inference, model routing, model split, and speculative decoding. 
However, existing evaluations are predominantly conducted in idealized datacenter settings, masking the practical challenges posed by heterogeneous hardware and stochastic wireless channels at 6G AI Edge.
This article presents the first unified, hardware-in-the-loop empirical
study that evaluates representative implementations of these paradigms
under practical network conditions.
Our stage-wise latency evaluation reveals that network dynamics can inflate Time-to-First-Token (TTFT) by up to $4.4\times$. Under normal network conditions, downlink hidden-state transmission accounts for approximately 491~ms of TTFT in the evaluated hidden-state-based Speculative Decoding implementation, while uplink transmission accounts for over 175~ms in Model Split.
Because no single paradigm universally dominates across all performance dimensions, we advocate for a context-aware adaptive orchestration framework that dynamically selects collaboration strategies based on network conditions, terminal capabilities, and task semantics, and outline a research roadmap toward 6G AI Edge deployment.
\end{abstract}
\begin{IEEEkeywords}
6G AI Edge, Large Language Models, Edge-Device Collaboration, Adaptive Orchestration.
\end{IEEEkeywords}

\section{Introduction}
\IEEEPARstart{T}{he} sixth-generation (6G) networks are poised to transform smartphones, autonomous robots, and connected vehicles from passive communication endpoints into proactive, embodied AI agents~\cite{cui2025overview}. 
Realizing this vision requires delivering generative AI capabilities (particularly, large language model (LLM) inference) directly at or near the user equipment.
However, a fundamental hardware mismatch persists: the rapid scaling of LLM parameters has vastly outpaced the compute, memory, and energy budgets of mobile terminals.
Standalone on-device inference remains prohibitive for complex multimodal tasks, necessitating external computational support to realize the 6G vision of ubiquitous intelligence.

\begin{figure*}[!t]
\centering

\subfloat[AI Cloud]{
    \includegraphics[width=0.285\textwidth]{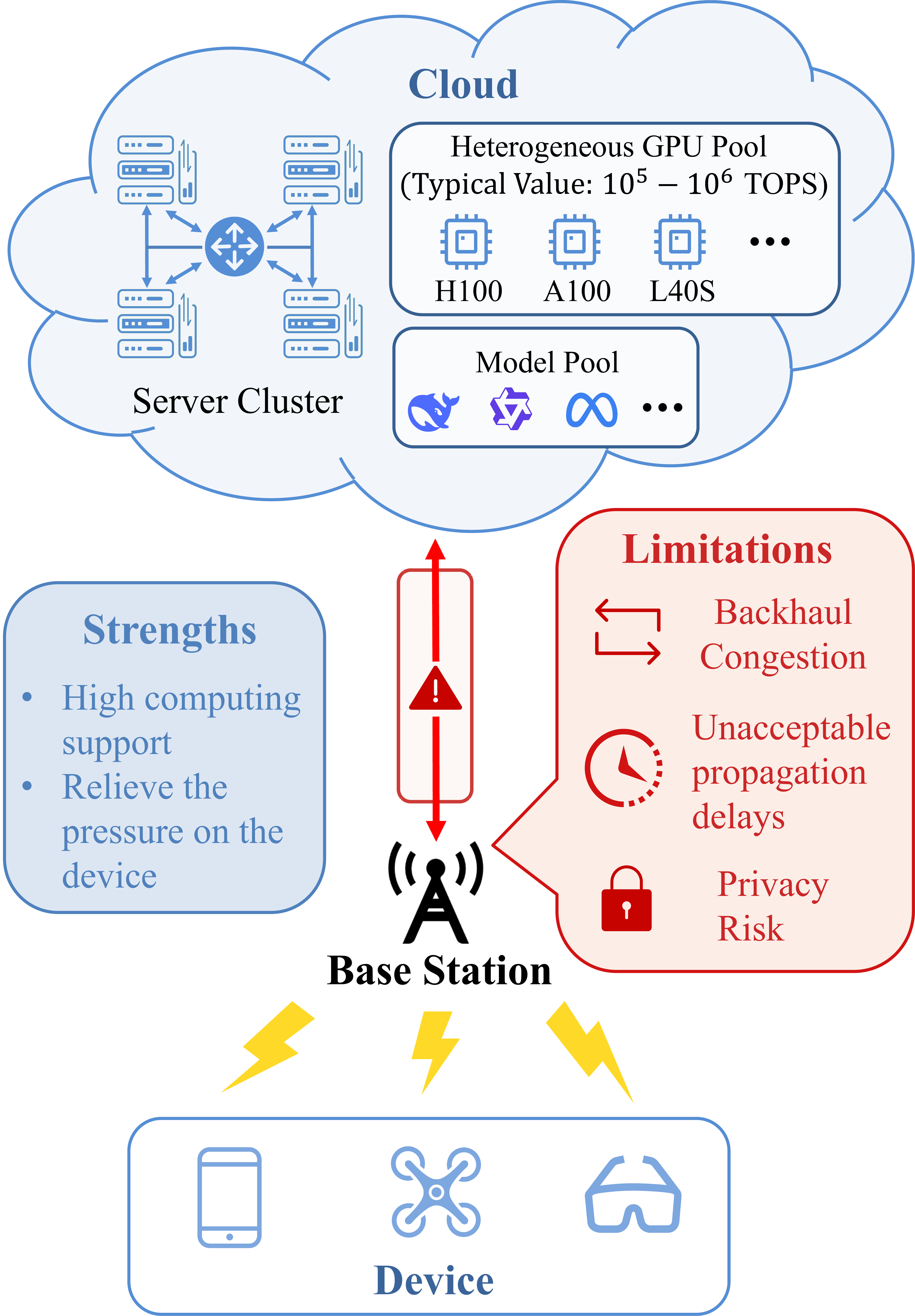}
    \label{fig:cloud_ai}
}
\hfill
\subfloat[6G AI Edge]{
    \includegraphics[width=0.67\textwidth]{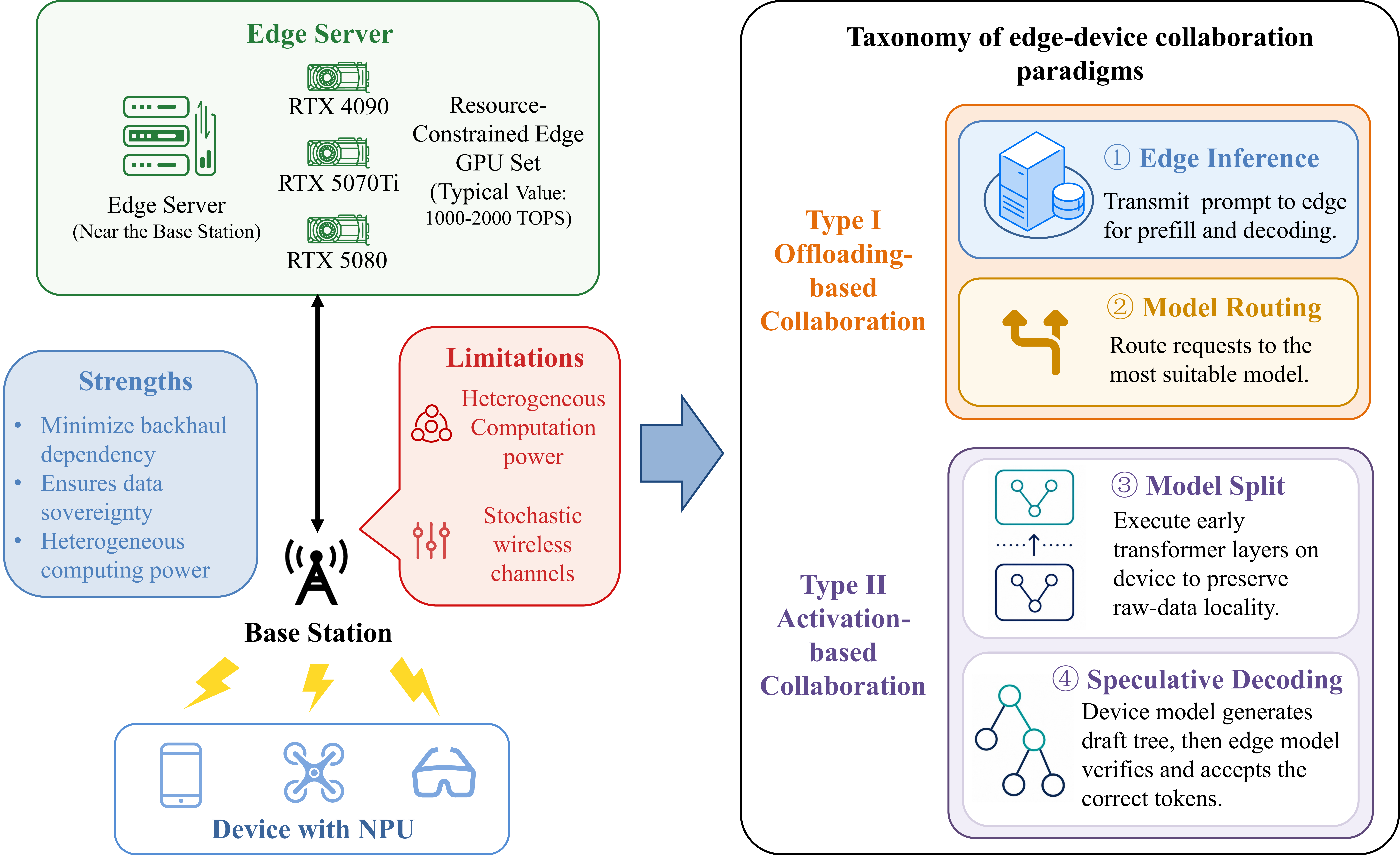}
    \label{fig:edge_ai}
}

\caption{The architectural shift from AI Cloud to AI Edge for 6G: (a) AI Cloud relies on distant datacenters, introducing severe backhaul bottlenecks and high latency. (b) AI Edge leverages proximate edge computing to enable edge-device collaborative LLM inference, which is categorized into a taxonomy of four paradigms.}
\label{fig:cloud_edge_ai}
\end{figure*}

Traditionally, this computational gap has been bridged by AI Cloud, where mobile terminals offload AI workloads to remote datacenters (Fig.~\ref{fig:cloud_ai}).
However, AI Cloud is fundamentally constrained by backhaul capacity and propagation latency: relying on distant servers introduces unacceptable delays, congestion, and privacy risks that violate the ultra-reliable, low-latency requirements of emerging 6G AI applications.
Consequently, 6G is decisively shifting toward the AI Edge era~\cite{yang2025aiedge} (Fig.~\ref{fig:edge_ai}),  which pushes LLM inference to network edge proximate to the user.
By minimizing backhaul dependency, preserving raw-data locality, and harnessing distributed heterogeneous resources, the AI Edge is emerging as the essential computing architecture for emerging 6G AI applications.

Edge-device collaborative LLM inference is a key enabler of this 6G AI Edge era~\cite{ma2025multitier}. 
As depicted in Fig.~\ref{fig:edge_ai}, it encompasses four paradigms organized into two categories.
\textit{Type I: Offloading-based Collaboration} comprises \textit{(1) Edge Inference}~\cite{yao-etal-2024-scalellm}--[\citen{srivatsa2025preble}], which executes the full model at the edge, and \textit{(2) Model Routing}~\cite{stripelis-etal-2024-tensoropera}--[\citen{chen2023frugalgpt}], which directs requests of different task types to suitable edge models.
\textit{Type II: Activation-based Collaboration} comprises \textit{(3) Model Split}~\cite{Ma2026SplitFrozen}--[\citen{ma2025fouriercompress}], which exchanges intermediate features while retaining raw inputs on the device, and \textit{(4) Speculative Decoding}~\cite{ning2025dssd}--[\citen{11551572}], which combines end-side drafting with edge-side verification.

While these mechanisms are well-studied in datacenter settings, the 6G AI Edge presents a fundamentally different operational reality characterized by two distinguishing features:
(1)~\textit{Heterogeneous computing capabilities}, ranging from resource-constrained mobile NPUs to edge GPUs, in contrast to homogeneous cloud clusters; and
(2)~\textit{Stochastic wireless channels} subject to bandwidth fluctuations, jitter, and packet loss, unlike the lossless terabit-speed inter-server links in datacenters.
This divergence raises a pivotal question:
\textit{Can these edge-device collaboration paradigms thrive under the 6G AI Edge conditions, and if not, how can we redesign them to ensure their viability?}
Answering this question demands moving beyond idealized datacenter settings to rigorous empirical evaluation under practical 6G AI Edge conditions, with a focus on LLM inference metrics, such as Time-to-First-Token (TTFT) and Time-Per-Output-Token (TPOT), rather than conventional throughput metrics.

To address this question, we present a comprehensive empirical study of edge-device collaborative LLM inference for 6G AI Edge. 
By deploying a unified, network-emulated hardware-in-the-loop testbed, we expose the hidden wireless bottlenecks that dictate real-world performance and outline a research roadmap toward 6G edge-device collaboration.
The key contributions are as follows:
\begin{itemize}
    \item We design a unified edge-device testbed pairing an NVIDIA Jetson Orin (device) with an RTX 4090 GPU server (edge) to capture heterogeneous computing capabilities.
    The testbed emulates channel dynamics (jitter, packet loss, and directional bandwidth limits) via kernel-level traffic shaping, enabling the first side-by-side empirical comparison of representative implementations of all four collaboration paradigms under the evaluated network conditions. We also assess sensitivity to end-side compute availability using two Jetson operating configurations.
    \item We perform a stage-wise latency and transmission breakdown, revealing that network-induced latency can inflate TTFT by up to $4.4\times$. Specifically, we expose severe tensor transmission bottlenecks: downlink hidden-state transmission accounts for approximately 491~ms of TTFT in the evaluated hidden-state-based Speculative Decoding implementation, while uplink transmission in Model Split consumes over 175~ms (nearly half of its total 367~ms TTFT).
    \item We find that no single paradigm universally dominates across all metrics; rather, optimal selection is strictly context-dependent. Accordingly, we argue that a context-aware and user-centric adaptive orchestration framework is critical for 6G AI Edge to dynamically select collaboration methods based on real-time network states, terminal capabilities, and task requirements.
\end{itemize}

The rest of this article is organized as follows. Section~II details the technical taxonomy of the collaboration paradigms. Section~III presents our testbed architecture and comparative performance analysis. Section~IV demonstrates the proposed adaptive orchestration framework and outlines future research directions, followed by the conclusion in Section~V.

\section{Preliminaries on Edge-Device Collaborative LLM Inference}

This section outlines the evolving standardization landscape of the 6G AI Edge, details the technical taxonomy of edge-device collaborative inference, and highlights the critical gaps that necessitate rethinking these paradigms for 6G.

\subsection{From Small-Model to Large-Model Collaboration}

AI standardization is extending support from individual models toward network-wide collaboration. 3GPP Release-18/19 address AI/ML model lifecycle management and model transfer, while SA2 explores AI/ML support for network automation, Quality of Service (QoS) optimization, and positioning. ETSI Multi-access Edge Computing (MEC) develops edge AI architectures, and Zero-touch network and Service Management (ZSM) addresses intent-driven automation. As generative AI expands, these efforts provide a broader context for studying large-model collaboration across devices, radio access networks, and the core.

\begingroup
\renewcommand{\thesubfigure}{\circledsubfiglabel}
\captionsetup[subfloat]{labelformat=simple,labelsep=space}

\begin{figure*}[!t]
\centering

\noindent
\begin{minipage}[c]{\rowdescw}
    \rowdesctext
    {Type I:}
    {Offloading-\\[-0.20em]based\\[-0.20em]Collaboration}
\end{minipage}%
\hfill
\begin{minipage}[c]{\rowcontentw}
\centering
    \paradigmpanel
    {Edge Inference}
    {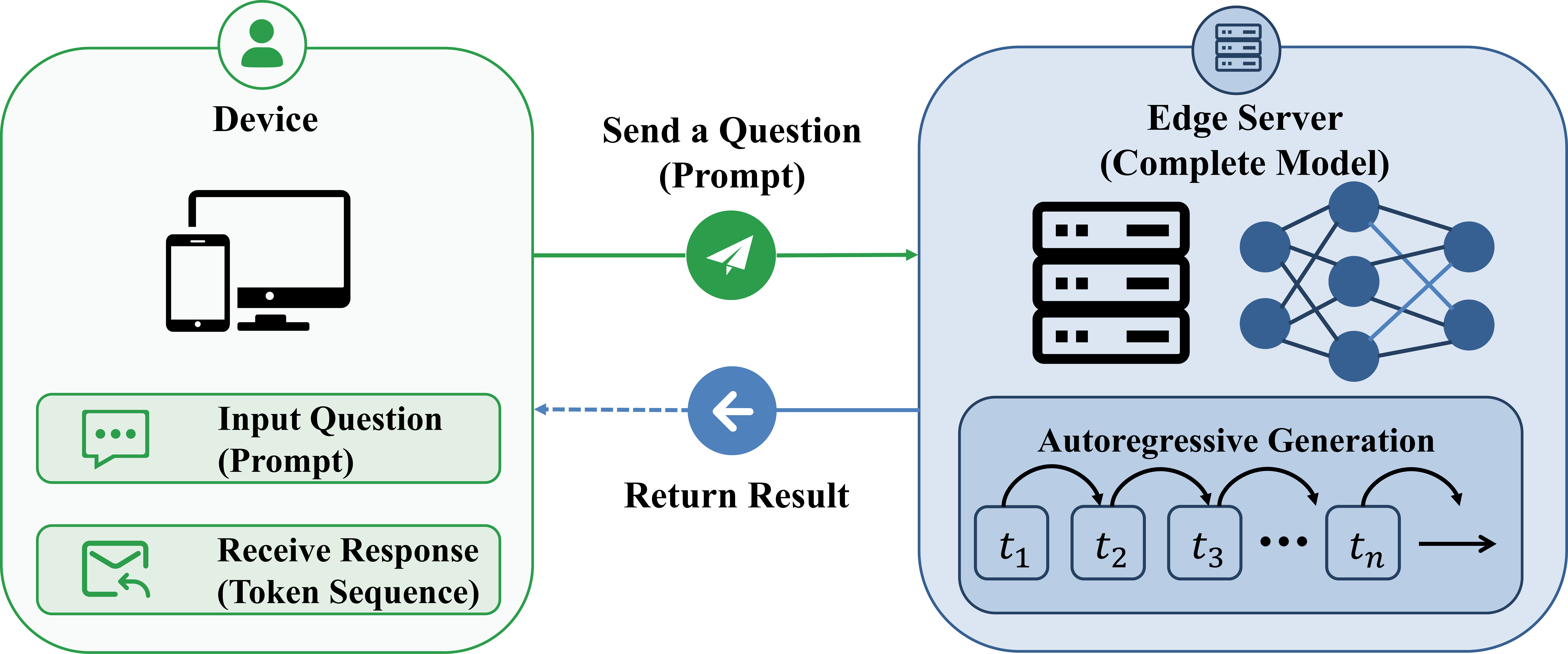}
    {ScaleLLM~\cite{yao-etal-2024-scalellm}, Preble~\cite{srivatsa2025preble}.}
    {Privacy exposure; TTFT inflated by wireless latency.} 
    {fig:edge_inference}
    \hfill
    \paradigmpanel
    {Model Routing}
    {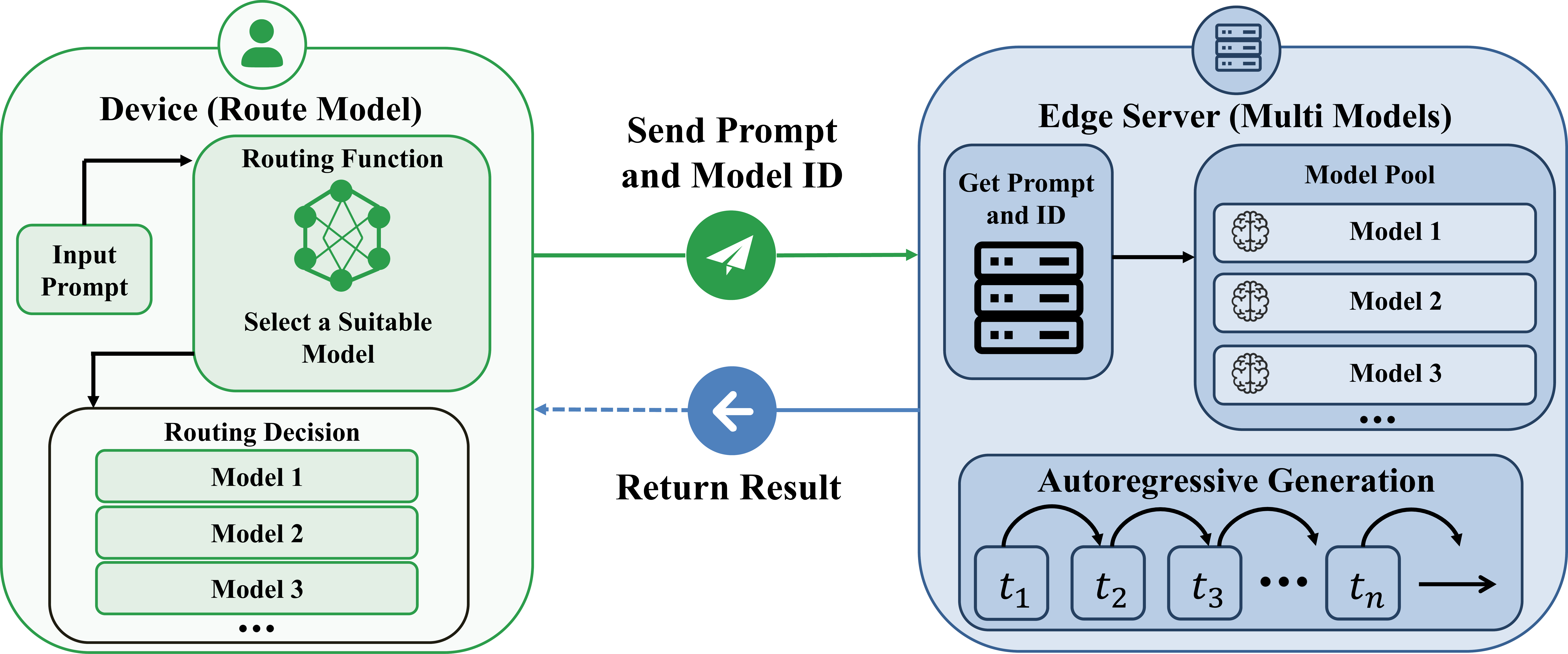}
    {TensorOpera Router~\cite{stripelis-etal-2024-tensoropera}, FrugalGPT~\cite{chen2023frugalgpt}.}
    {Misrouting triggers fallbacks, causing latency penalties.} 
    {fig:model_routing}
\end{minipage}

\rowdashsep

\noindent
\begin{minipage}[c]{\rowdescw}
    \rowdesctext
    {Type II:}
    {Activation-\\[-0.20em]based\\[-0.20em]Collaboration}
\end{minipage}%
\hfill
\begin{minipage}[c]{\rowcontentw}
\centering
    \paradigmpanel
    {Model Split}
    {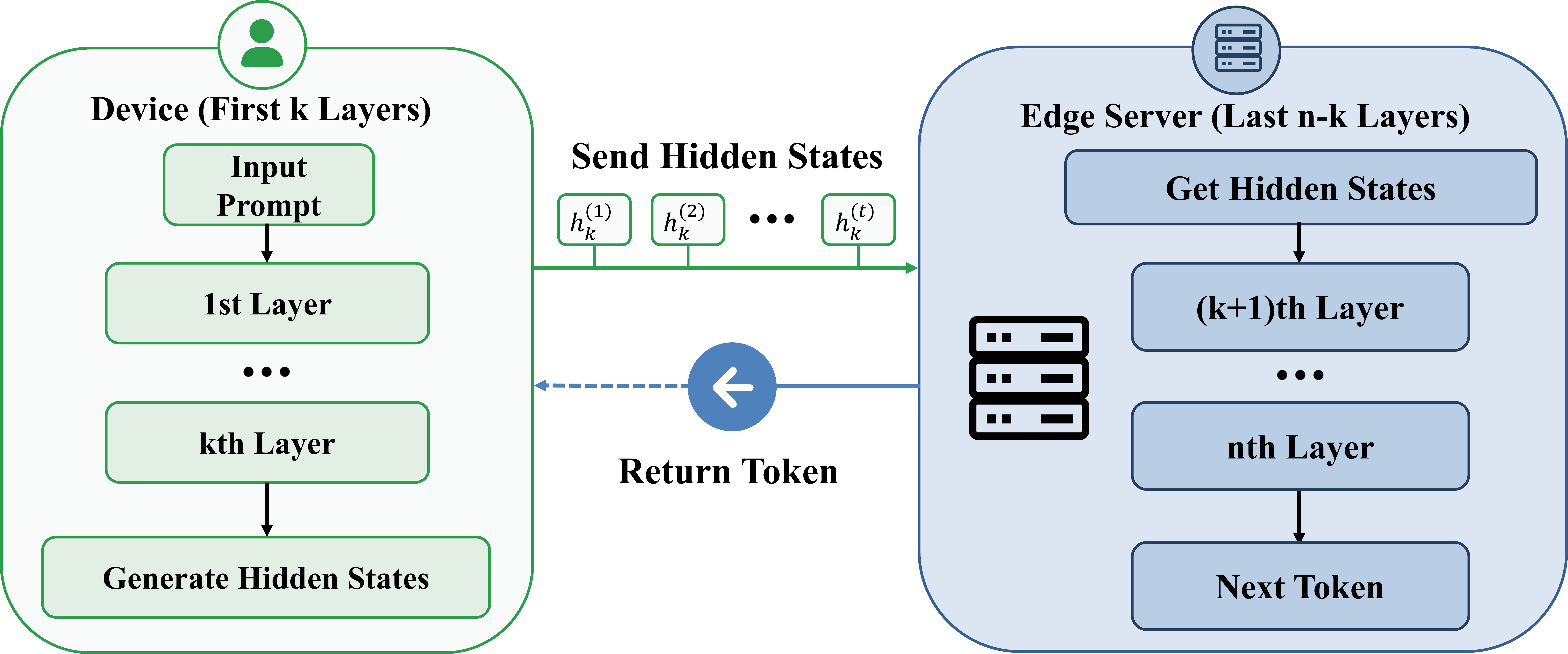}
    {SplitFrozen~\cite{Ma2026SplitFrozen}, FourierCompress~\cite{ma2025fouriercompress}.}
    {Uplink tensor bottlenecks inflate TTFT; vulnerable to packet loss.} 
    {fig:model_split}
    \hfill
    \paradigmpanel
    {Speculative Decoding}
    {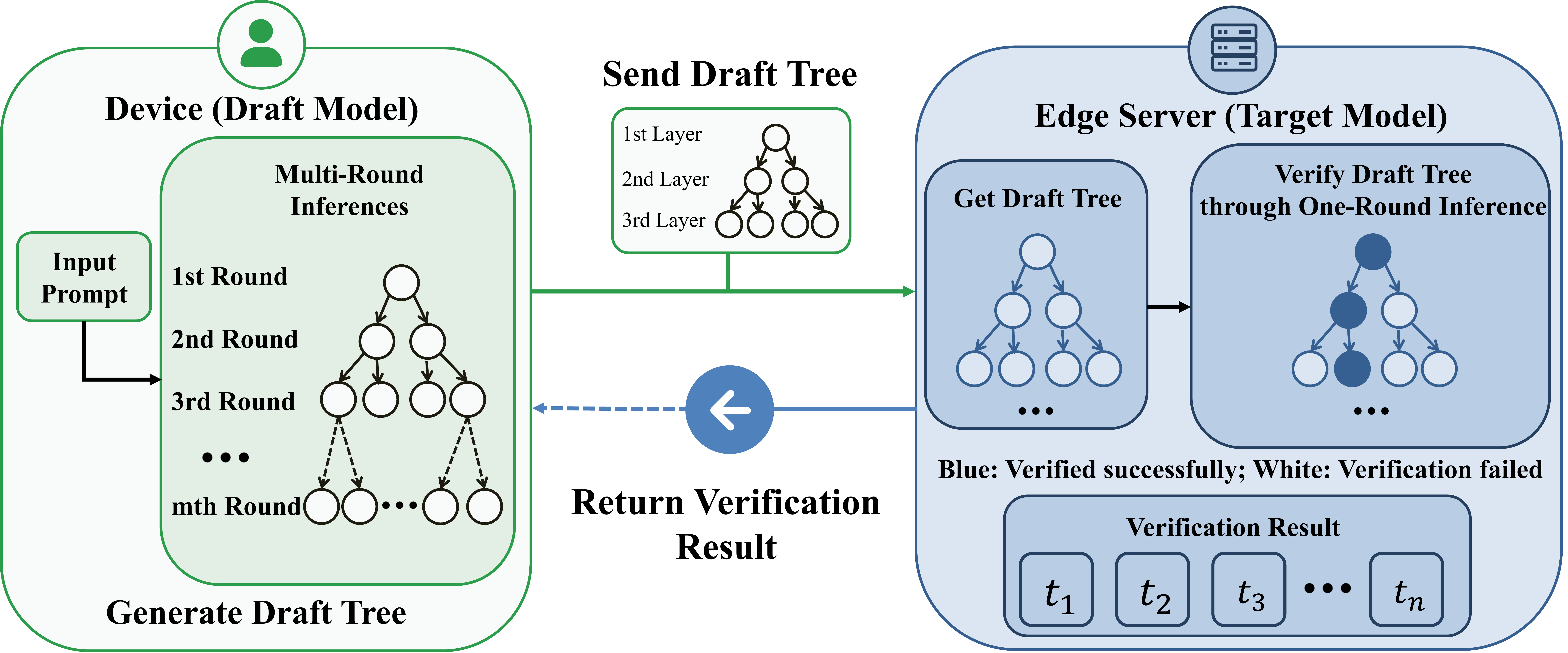}
    {DSSD~\cite{ning2025dssd}, CU-HLM~\cite{11551572}.}
    {Heavy downlink transfers negate TPOT speedups.} 
    {fig:speculative_decoding}
\end{minipage}

\caption{Taxonomy and technical workflows of edge-device collaborative LLM inference, illustrating their execution workflows, representative works, and limitations.}
\label{fig:framework}
\end{figure*}
\endgroup

\subsection{Taxonomy of Edge-Device Collaborative LLM Inference}
In the 6G AI Edge landscape, edge-device collaboration paradigms address stringent terminal hardware constraints (encompassing computation, memory, and energy) through strategic workload partitioning. As illustrated in Fig.~\ref{fig:framework}, this landscape is structured around four distinct paradigms, categorized into offloading-based and activation-based collaboration.

\subsubsection{Edge Inference}
This baseline paradigm treats the terminal as a thin client, offloading the computational burden entirely to the edge. 
For instance, ScaleLLM~\cite{yao-etal-2024-scalellm} upgrades engine gateways for end-to-end efficiency, and Preble~\cite{srivatsa2025preble} employs heuristic scheduling to balance cache affinity. 
However, these solutions inherently assume lossless, high-speed internal interconnects for datacenter settings.

\textit{Workflow:} The execution follows a sequential pipeline. \textit{Step 1:} The terminal transmits the lightweight user prompt over the uplink channel. 
\textit{Step 2:} The edge server executes the prefill phase (populating the KV cache) and the autoregressive decode phase. \textit{Step 3:} The generated tokens are streamed back to the terminal via the downlink for user presentation.

\subsubsection{Model Routing}
Model routing identifies the task type of each incoming request and directs it to an appropriate edge model according to the model's capabilities for that type of request, while considering inference quality, latency, and cost. Frameworks like TensorOpera Router~\cite{stripelis-etal-2024-tensoropera} learn prompt-to-expert classifiers from expert-response metrics and dynamically route each incoming query to the predicted most suitable LLM expert, and FrugalGPT~\cite{chen2023frugalgpt} uses prompt adaptation and adaptive cascades to drastically cut inference costs.

\textit{Workflow:} Intelligence is distributed through a request-classification and model-selection pipeline. \textit{Step 1:} The terminal identifies the task type of the incoming request, such as scientific question answering, everyday practical reasoning, programming, or economics. \textit{Step 2:} The router matches the identified request type to a suitable model in the edge model pool. \textit{Step 3:} The selected model processes the request and streams the output back to the terminal.

\subsubsection{Model Split}
Model Split retains raw inputs on the device and transmits intermediate hidden representations to the edge, avoiding direct transmission of raw inputs. However, these representations may still reveal or enable reconstruction of input content; therefore, raw-data locality does not guarantee input confidentiality. Recent wireless-focused studies address the resulting communication overhead through innovations like SplitFrozen~\cite{Ma2026SplitFrozen}, which utilizes edge-side LoRA for non-IID data, and FourierCompress~\cite{ma2025fouriercompress}, which applies layer-aware spectral compression to intermediate tensors.


\textit{Workflow:} The forward pass is physically bifurcated across the network. \textit{Step 1:} The terminal executes the token embedding and the first $k$ transformer layers, generating an intermediate hidden-state tensor. \textit{Step 2:} This tensor is transmitted over the uplink to the edge. \textit{Step 3:} The edge server receives the hidden states, updates its KV cache, and executes the remaining layers and the language model head to generate the output tokens.

\subsubsection{Speculative Decoding}
This paradigm accelerates generation by leveraging a lightweight on-device draft model to propose candidate tokens, which are then verified by a larger edge model. Recent advancements include Distributed Split Speculative Decoding~\cite{ning2025dssd} to reduce communication, and uncertainty-aware opportunistic transmission~\cite{11551572} to compress payloads. Despite these algorithmic gains, the wireless transmission of verification payloads remains a severe bottleneck. Speculative decoding encompasses multiple algorithmic designs with
different draft-generation mechanisms and communication requirements.
For the empirical study in Section III, we instantiate this paradigm
using EAGLE-3~\cite{li2025eagle3}, a representative and competitive
modern speculative decoding method. EAGLE-3 directly predicts draft
tokens using a lightweight draft model conditioned on fused low-,
middle-, and high-level features from the target model, and employs
a dynamic draft tree for parallel verification. When the draft model is placed on the device and the target model is
placed at the edge, as in our testbed, the target-derived features
required by EAGLE-3 must be synchronized between the two endpoints.
Therefore, the communication overhead measured in this study is
specific to our hidden-state-based edge-device deployment and should not
be interpreted as an inherent requirement of all speculative decoding
methods.

\textit{Workflow:} In the evaluated hidden-state-based deployment, the terminal first sends the input token identifiers to the edge for prefill. The edge returns the first generated token and target-model hidden states to initialize the end-side draft model. The decoding stage then proceeds as follows. \textit{Step 1:} The terminal's draft model generates a draft tree of candidate tokens. \textit{Step 2:} The draft tree and its associated verification inputs are transmitted uplink to the edge. \textit{Step 3:} The edge's large target model performs a single forward pass to verify the candidates in parallel. \textit{Step 4:} The edge returns the verified tokens and the target-model hidden states needed for subsequent draft generation. This exchange repeats during decoding.

\subsection{6G AI Edge Reality: Rethinking Collaboration Mechanisms}
Heterogeneous hardware and wireless impairments challenge collaboration mechanisms designed for datacenters. Large hidden-state transfers in the evaluated hidden-state-based Speculative Decoding implementation and Model Split can create downlink and uplink bottlenecks, respectively, while wireless propagation delays and jitter can inflate TTFT for edge inference. These challenges motivate the unified, hardware-in-the-loop evaluation in the following section, using TTFT and TPOT to assess performance under controlled network conditions.

\begingroup
\renewcommand{\thesubfigure}{\figthreecircledsubfiglabel}
\captionsetup[subfloat]{labelformat=simple,labelsep=space}

\begin{figure*}[!t]
    \centering

    \includegraphics[width=0.98\textwidth]{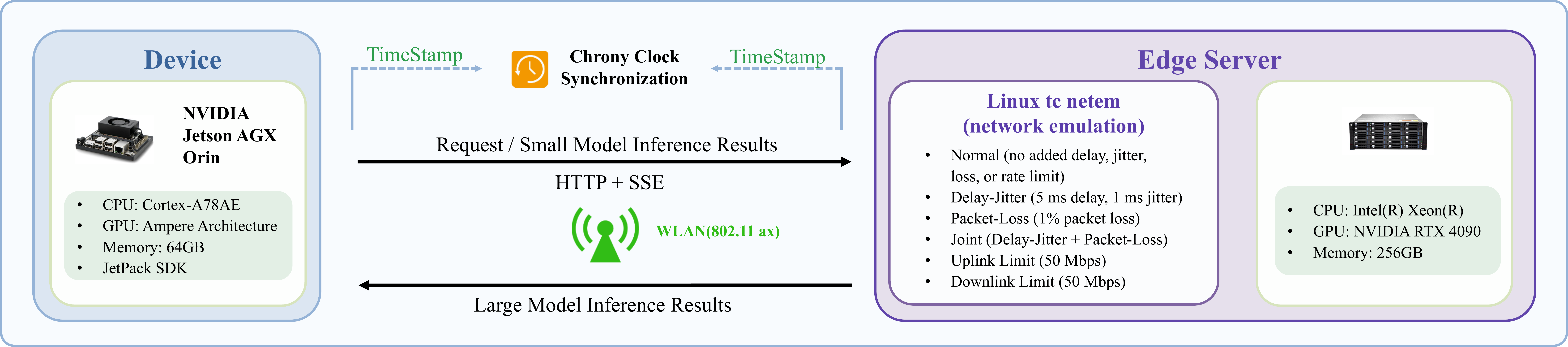}

    {\small System hardware and software architecture}

    \vspace{0.15cm}

    \setlength{\tabcolsep}{4pt}
    \begin{tabular}{@{}cc@{}}
        \subfloat[Edge Inference]{
            \includegraphics[width=0.48\textwidth]{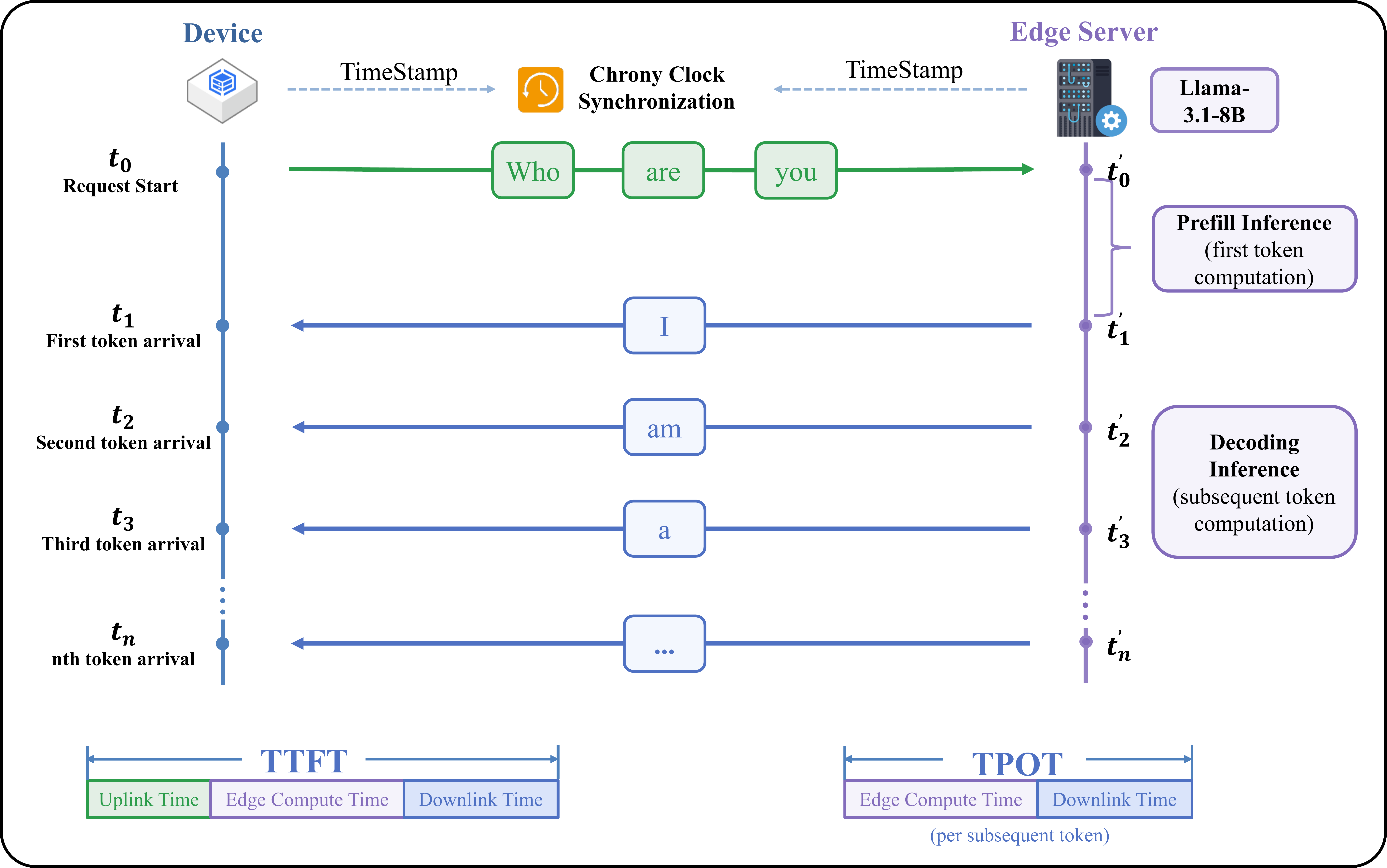}
            \label{fig:fig3_edge_inference}
        }
        &
        \subfloat[Model Routing]{
            \includegraphics[width=0.48\textwidth]{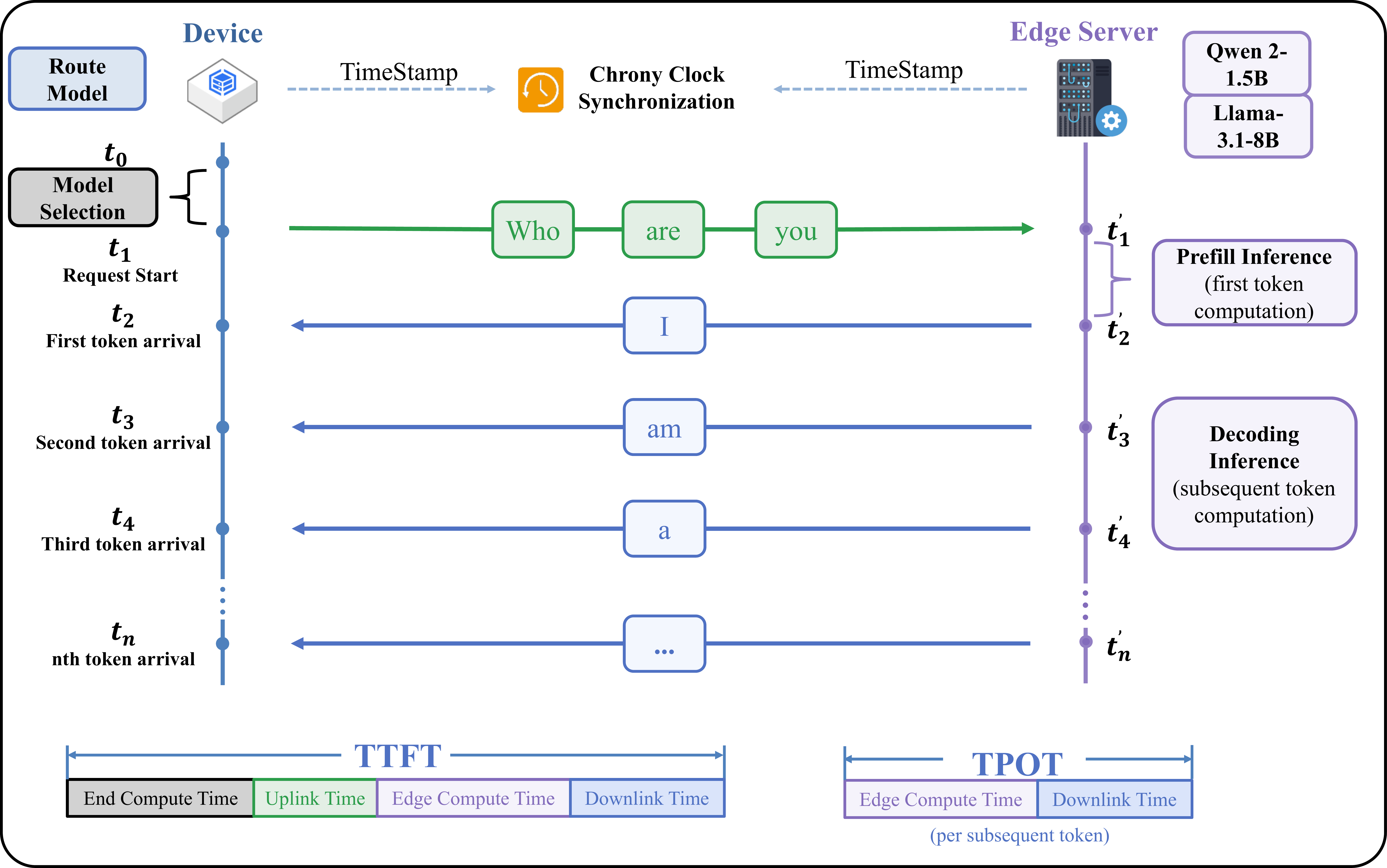}
            \label{fig:fig3_model_routing}
        }
    \end{tabular}

    \vspace{0.15cm}

    \setlength{\tabcolsep}{4pt}
    \begin{tabular}{@{}cc@{}}
        \subfloat[Model Split]{
            \includegraphics[width=0.48\textwidth]{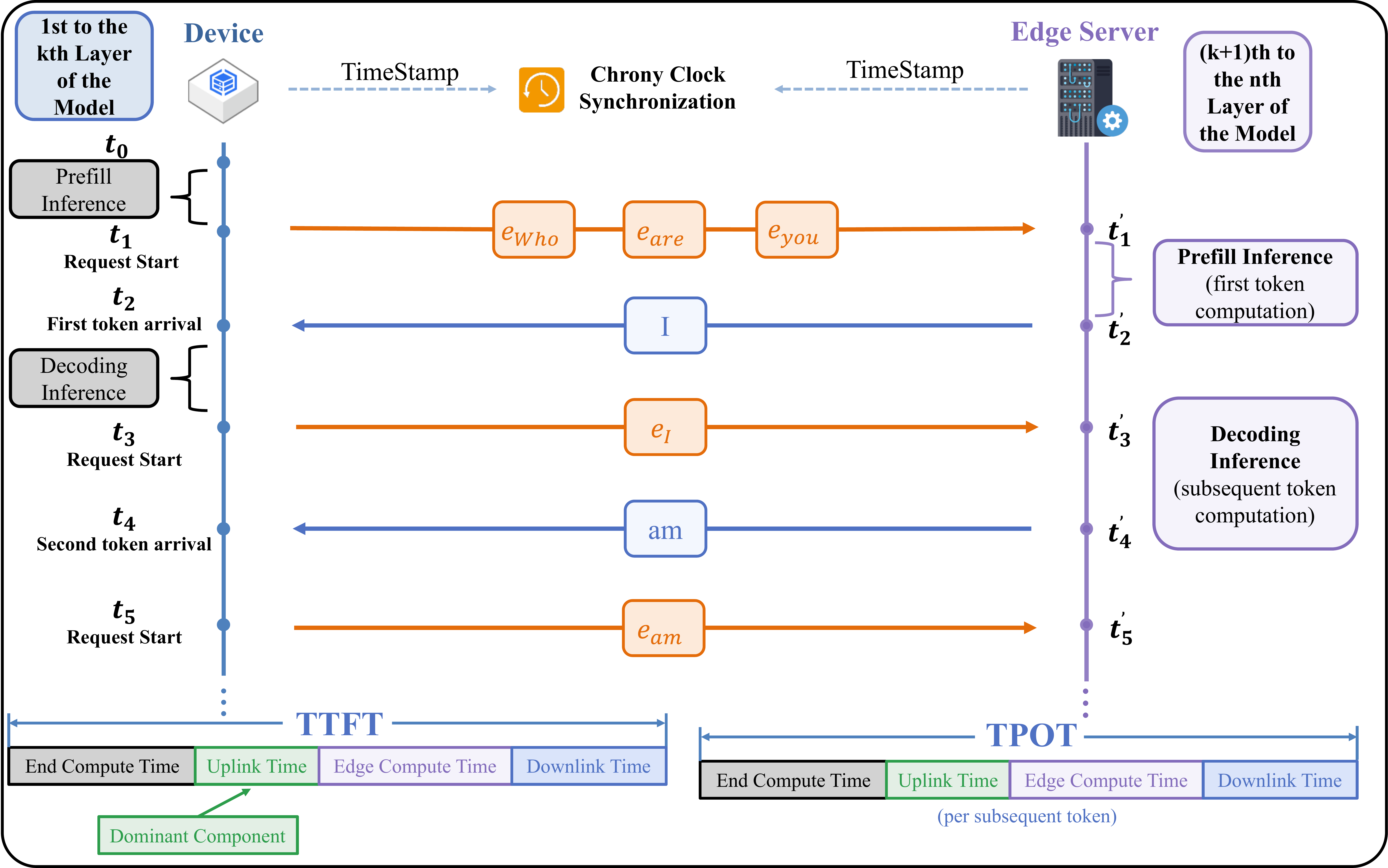}
            \label{fig:fig3_model_split}
        }
        &
        \subfloat[Speculative Decoding]{
            \includegraphics[width=0.48\textwidth]{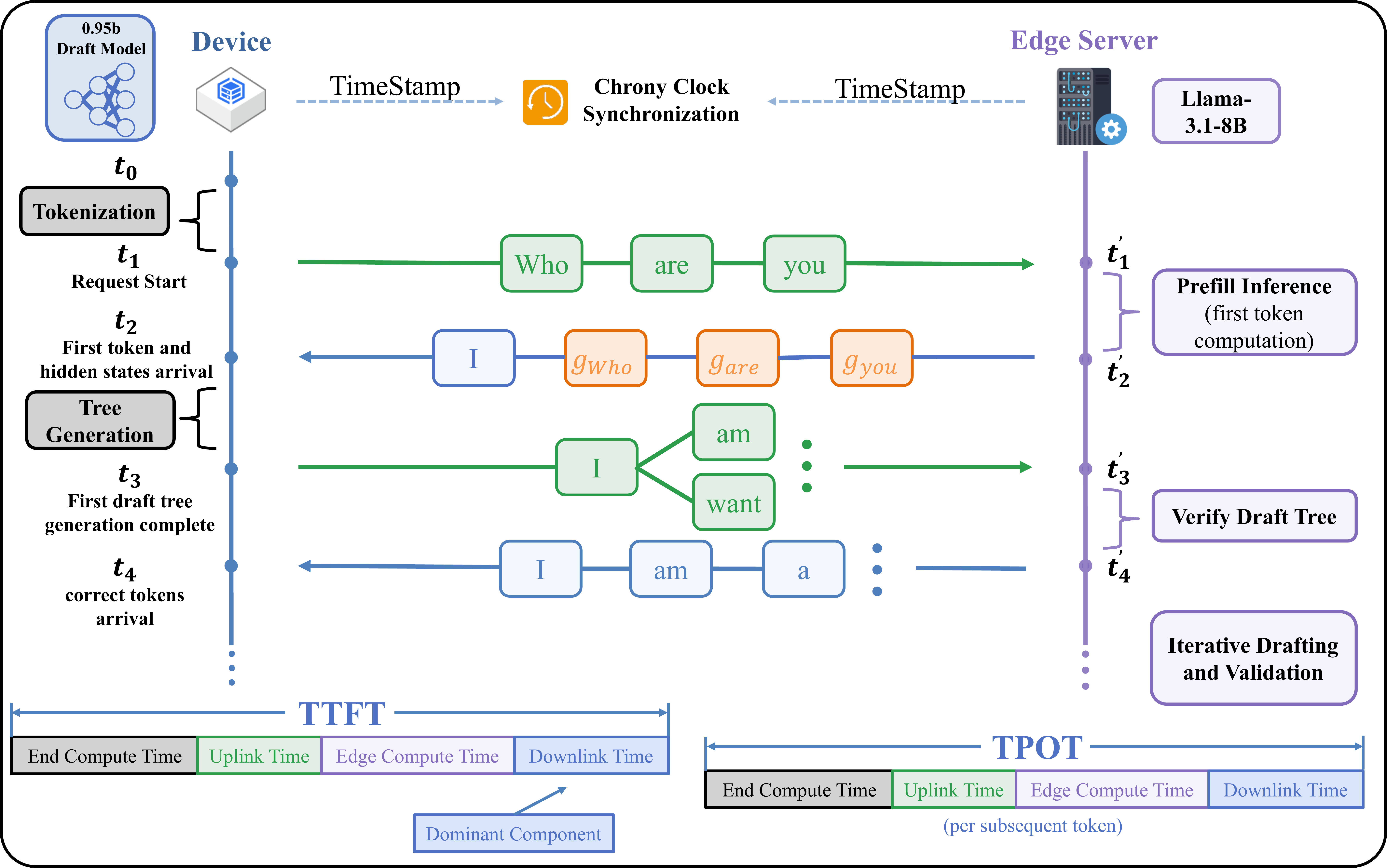}
            \label{fig:fig3_speculative_decoding}
        }
    \end{tabular}

    \caption{Unified Cloud-Edge-Device Collaborative Inference Testbed. The setup connects a Jetson Orin terminal to an RTX 4090 edge server via a local area network (LAN). The software stack is implemented using the Hugging Face Transformers library, enabling flexible model execution of four edge-device collaboration paradigms.}
    \label{fig:hardware_and_inference}
\end{figure*}
\endgroup

\section{Testbed Design and Case Studies}
In this section, we constructed a unified testbed that replicates the heterogeneous computing power and wireless conditions of 6G AI Edge.
This section details our testbed design, presents the performance analysis, and synthesizes the critical trade-offs of edge-device collaboration.

\subsection{Testbed Architecture and Experimental Configuration}
A core challenge in evaluating edge-device collaboration is integrating four distinct paradigms into a single, reproducible platform while accurately emulating wireless impairments. Our testbed bridges this gap through a tightly integrated hardware-software stack.

\textit{1) Heterogeneous Hardware Stack.} As illustrated in Fig.~\ref{fig:hardware_and_inference}, the user device is represented by an NVIDIA Jetson AGX Orin (64GB), capturing the memory-constrained, NPU-limited reality of 6G intelligent devices.
The edge server is powered by an NVIDIA RTX 4090 GPU server, emulating a high-performance MEC server. 
Rather than relying on lossless internal buses, the device and edge server communicate over a network link that is actively shaped to mimic 6G wireless characteristics (as in the following). To characterize the sensitivity of collaborative inference to end-side compute heterogeneity, we evaluate two operating configurations on the same NVIDIA Jetson AGX Orin device. The normal configuration operates in the official \texttt{MODE\_30W} profile (mode ID~2), with eight online CPU cores at up to 1.728 GHz, four GPU TPCs at up to 612 MHz, and an EMC frequency of up to 3.2 GHz. The resource-constrained configuration limits the same device to two online CPU cores at up to 1.114 GHz, one GPU TPC at up to 408 MHz, and an EMC frequency of up to 2.133 GHz. Both configurations are evaluated under the normal network profile using the same fixed mixed workload and measurement procedure.


\textit{2) Network Emulation via \texttt{tc-netem}.}
We use Linux Traffic Control (\texttt{tc-netem}) to impose
controlled packet-level impairments and directional rate limits on the traffic exchanged
between the device and the edge server. Under natural wireless conditions without artificial
traffic shaping, we measured the baseline end-to-end throughput using \texttt{iperf3} with a
single TCP stream. We conducted 10 runs per direction, each lasting 30~s, with the first
3~s excluded as warm-up. Uplink and downlink throughputs were
$248.43 \pm 6.64$~Mbps and $257.65 \pm 10.33$~Mbps, respectively
(mean $\pm$ sample standard deviation across runs), with corresponding medians of
248.88 and 259.87~Mbps. Six network profiles are evaluated:
(1) \textit{Normal}, with no additional delay, jitter, packet loss, or rate limit imposed
by \texttt{tc-netem};
(2) \textit{Delay-Jitter}, with a 5~ms delay and 1~ms jitter applied in each direction;
(3) \textit{Packet-Loss}, with 1\% random packet loss applied in each direction;
(4) \textit{Joint}, combining the Delay-Jitter and Packet-Loss settings;
(5) \textit{Uplink-Limited}, with the device-to-edge rate capped at 50~Mbps and no
additional rate limit on the downlink; and
(6) \textit{Downlink-Limited}, with the edge-to-device rate capped at 50~Mbps and no
additional rate limit on the uplink. We acknowledge that the \texttt{tc-netem} settings
provide a simplified approximation of wireless conditions, where HARQ retransmissions and time-varying capacities are not fully captured. Real wireless conditions may therefore be even more adverse than those emulated in this paper.


\textit{3) Model Configurations.}
Across the four paradigms, Edge Inference uses Llama-3.1-8B-Instruct.
Model Routing employs Qwen2-1.5B-Instruct and
Llama-3.1-8B-Instruct as the candidate models. For the evaluated
hidden-state-based Speculative Decoding implementation,
Llama-3.1-8B-Instruct is deployed at the edge, while the corresponding
EAGLE-3 draft model is deployed on the device. For Model Split,
Llama-3.1-8B-Instruct is partitioned after the first Transformer layer ($k=1$):
the device executes the embedding layer and the first Transformer layer,
whereas the edge executes the remaining 31 Transformer layers and the
language-model head. All models used in the four paradigms are loaded
in bfloat16 precision.

\textit{4) Metric Decomposition and Measurement.}
Performance is evaluated through two user-centric KPIs:
TTFT (which dictates initial system responsiveness), and TPOT
(which measures sustained generation efficiency).
Latency experiments use a frozen workload of 10 English
multiple-choice queries, with two each from ARC-Easy, PIQA,
SciQ, ARC-Challenge, and MMLU-Pro. By content, these queries cover
natural science (six), everyday practical reasoning (two), programming
(one), and economics (one). Each model receives a 200-token full
input and generates exactly 64 tokens. The same queries and
order are used across paradigms and network conditions.
To isolate communication-compute trade-offs, both metrics
are decomposed into four stages: on-device execution, uplink
transmission, edge execution, and downlink transmission.
Precise stage-wise measurement is achieved via Chrony-based
clock synchronization ($<100\,\mu$s accuracy) and FastAPI
with Server-Sent Events (SSE) for token-level streaming
telemetry. Our Model Routing heuristic scores request keywords, length, and
multi-step markers, assigning six requests to Qwen2-1.5B-Instruct and
four to Llama-3.1-8B-Instruct. In our implementation, request features are used as heuristic routing signals; the content categories describe the workload rather than define a fixed category-to-model mapping. To quantify
the quality--latency trade-off between these candidates, we evaluate
both models on ARC-Easy, PIQA, and SciQ, using 500 questions from
each dataset with zero-shot prompting and greedy decoding. For Speculative Decoding, the mean accepted
draft tokens $\bar a$ and mean selected path depth $\bar d$
exclude the root and are averaged over decoding rounds
within each request, then across retained requests.
The acceptance rate is defined as $\alpha=\bar a/\bar d$.

\begin{table*}[!t]
\centering
\caption{Comprehensive Latency and Transmission Breakdown of Edge-Device Collaborative Inference}
\label{tab:comprehensive_latency}
\footnotesize
\setlength{\tabcolsep}{3pt}
\renewcommand{\arraystretch}{1.1}

\begin{tabular*}{\textwidth}{@{\extracolsep{\fill}} l l c c c c c @{}}
\toprule
\multicolumn{7}{@{}l}{\textbf{Panel A: Stage-wise Latency for a Fixed-Length Mixed Workload under Different Network Conditions (TTFT / TPOT in ms)}} \\
\midrule
\textbf{Network Profile} & \textbf{Paradigm} & \textbf{Device Stage} & \textbf{Uplink} & \textbf{Edge Stage} & \textbf{Downlink} & \textbf{Total} \\
\midrule
\multirow{4}{*}{Normal}
 & Edge Inference & 2.12 / 0.00 & 3.62 / 0.00 & 65.50 / 38.70 & 1.89 / 2.50 & \textbf{73.13 / 41.20} \\
 & Model Routing & 3.74 / 0.00 & 4.63 / 0.00 & 49.21 / 29.87 & 3.53 / 3.24 & 61.11 / 33.12 \\
 & Model Split & 107.82 / 33.25 & 175.19 / 7.24 & 78.70 / 59.39 & 5.40 / 3.85 & 367.10 / 103.73 \\
 & Speculative Decoding & 1.77 / 28.45 & 9.32 / 4.06 & 110.93 / 19.62 & 491.37 / 11.49 & 613.39 / 63.61 \\
\midrule
\multirow{4}{*}{\makecell[l]{Delay-Jitter \\ (5 ms, 1 ms)}}
 & Edge Inference & 1.95 / 0.00 & 9.71 / 0.00 & 64.34 / 37.84 & 7.41 / 7.50 & 83.41 / 45.34 \\
 & Model Routing & 3.74 / 0.00 & 10.38 / 0.00 & 46.98 / 28.94 & 8.05 / 7.82 & 69.16 / 36.76 \\
 & Model Split & 111.51 / 44.48 & 271.39 / 24.20 & 81.28 / 59.29 & 10.73 / 9.54 & 474.90 / 137.51 \\
 & Speculative Decoding & 1.66 / 28.58 & 22.46 / 9.71 & 119.24 / 21.94 & 597.43 / 12.80 & 740.78 / 73.02 \\
\midrule
\multirow{4}{*}{\makecell[l]{Packet-Loss \\ (1\% loss)}}
 & Edge Inference & 1.95 / 0.00 & 4.27 / 0.00 & 65.87 / 38.52 & 2.34 / 3.04 & 74.42 / 41.56 \\
 & Model Routing & 3.64 / 0.00 & 4.99 / 0.00 & 44.33 / 28.64 & 3.54 / 3.55 & 56.50 / 32.20 \\
 & Model Split & 110.02 / 36.87 & 345.60 / 29.04 & 84.40 / 59.81 & 6.35 / 4.86 & 546.37 / 130.58 \\
 & Speculative Decoding & 1.60 / 29.50 & 33.90 / 10.09 & 114.20 / 22.41 & 1337.58 / 6.42 & 1487.28 / 68.42 \\
\midrule
\multirow{4}{*}{\makecell[l]{Joint \\ (Jitter + Loss)}}
 & Edge Inference & 1.88 / 0.00 & 9.33 / 0.00 & 66.18 / 39.32 & 7.14 / 9.22 & 84.53 / 48.54 \\
 & Model Routing & 4.07 / 0.00 & 9.68 / 0.00 & 48.75 / 28.68 & 7.46 / 8.59 & 69.95 / 37.27 \\
 & Model Split & 105.86 / 45.65 & 1070.00 / 50.51 & 88.35 / 60.68 & 11.70 / 9.98 & 1275.92 / 166.81 \\
 & Speculative Decoding & 1.87 / 33.12 & 34.04 / 18.66 & 123.95 / 24.12 & 2532.40 / 14.80 & 2692.26 / 90.69 \\
\midrule
\multirow{4}{*}{\makecell[l]{Uplink Limit \\ (50 Mbps)}}
 & Edge Inference & 1.83 / 0.00 & 4.76 / 0.00 & 65.13 / 38.73 & 2.89 / 3.04 & 74.61 / 41.77 \\
 & Model Routing & 3.37 / 0.00 & 5.08 / 0.00 & 48.11 / 29.28 & 4.18 / 4.78 & 60.74 / 34.06 \\
 & Model Split & 106.29 / 33.23 & 408.46 / 7.95 & 84.20 / 59.39 & 6.20 / 3.65 & 605.15 / 104.23 \\
 & Speculative Decoding & 1.59 / 29.41 & 13.82 / 4.97 & 112.57 / 20.35 & 503.59 / 11.78 & 631.56 / 66.51 \\
\midrule
\multirow{4}{*}{\makecell[l]{Downlink Limit \\ (50 Mbps)}}
 & Edge Inference & 1.69 / 0.00 & 4.47 / 0.00 & 65.14 / 38.87 & 1.74 / 3.56 & 73.04 / 42.43 \\
 & Model Routing & 4.01 / 0.00 & 5.04 / 0.00 & 45.76 / 28.88 & 3.05 / 3.39 & 57.87 / 32.27 \\
 & Model Split & 109.69 / 31.97 & 158.79 / 7.19 & 78.18 / 59.36 & 5.32 / 3.58 & 351.99 / 102.10 \\
 & Speculative Decoding & 1.76 / 30.85 & 19.41 / 4.44 & 111.74 / 20.98 & 1618.77 / 12.70 & 1751.69 / 68.97 \\
\bottomrule
\end{tabular*}

\vspace{2mm}



\begin{tabular*}{\textwidth}{@{\extracolsep{\fill}} l c c c c c @{}}
\toprule
\multicolumn{6}{@{}l}{\textbf{Panel B: Model Routing Latency by Selected Model (Normal Network, TTFT / TPOT in ms)}} \\
\midrule
\textbf{Selected Model} & \textbf{Device Stage} & \textbf{Uplink} & \textbf{Edge Stage} & \textbf{Downlink} & \textbf{Total} \\
\midrule
Large (Llama-3.1-8B) & 4.08 / 0.00 & 4.59 / 0.00 & 62.73 / 37.56 & 3.26 / 3.39 & 74.66 / 40.95 \\
Small (Qwen2-1.5B)  & 3.47 / 0.00 & 4.66 / 0.00 & 38.39 / 23.72 & 3.74 / 3.13 & \textbf{50.27 / 26.85} \\
\bottomrule
\end{tabular*}

\vspace{2mm}


\begin{tabular*}{\textwidth}{@{\extracolsep{\fill}} l c c c c l @{}}
\toprule
\multicolumn{6}{@{}l}{\textbf{Panel C: Transmission Overhead by Paradigm (Normal Network, KB)}} \\
\midrule
\textbf{Paradigm} & \textbf{Prefill UL} & \textbf{Prefill DL} & \textbf{Decode UL} & \textbf{Decode DL} & \textbf{Bottleneck Remark} \\
\midrule
Edge Inference
& 0.40
& 0.05
& -
& 0.05
& Minimal bandwidth footprint \\

Model Routing
& 0.55
& 0.23
& -
& 0.23
& Minimal bandwidth footprint \\

Model Split
& \textbf{2323.85}
& 1.79
& 11.79
& 1.79
& Heavy UL feature tensor transfer \\

Speculative Decoding
& 2.28
& \textbf{6401.71}
& 22.84 / 23.48
& 114.30 / 194.23
& Massive DL hidden-state transfer \\
\bottomrule
\end{tabular*}

\vspace{2mm}

\begin{tabular*}{\textwidth}{
@{\extracolsep{\fill}} l l c c c c c @{}}
\toprule
\multicolumn{7}{@{}l}{
\textbf{Panel D: Stage-wise Latency under Different Device Configurations
(Normal Network, TTFT / TPOT in ms)}
} \\
\midrule
\textbf{Device Configuration}
& \textbf{Paradigm}
& \textbf{Device Stage}
& \textbf{Uplink}
& \textbf{Edge Stage}
& \textbf{Downlink}
& \textbf{Total} \\
\midrule

\multirow{4}{*}{
\makecell[l]{Normal\\(CPU8/TPC4)}
}
& Edge Inference
& 2.12 / 0.00
& 3.62 / 0.00
& 65.50 / 38.70
& 1.89 / 2.50
& 73.13 / 41.20 \\

& Model Routing
& 3.74 / 0.00
& 4.63 / 0.00
& 49.21 / 29.87
& 3.53 / 3.24
& 61.11 / 33.12 \\

& Model Split
& 107.82 / 33.25
& 175.19 / 7.24
& 78.70 / 59.39
& 5.40 / 3.85
& 367.10 / 103.73 \\

& Speculative Decoding
& 1.77 / 28.45
& 9.32 / 4.06
& 110.93 / 19.62
& 491.37 / 11.49
& 613.39 / 63.61 \\

\midrule

\multirow{4}{*}{
\makecell[l]{Resource-Constrained\\(CPU2/TPC1)}
}
& Edge Inference
& 1.09 / 0.00
& 2.96 / 0.00
& 65.78 / 39.15
& 1.32 / 1.56
& 71.15 / 40.70 \\

& Model Routing
& 2.47 / 0.00
& 4.03 / 0.00
& 48.07 / 29.50
& 3.51 / 3.30
& 58.09 / 32.79 \\

& Model Split
& \textbf{143.36 / 63.83}
& 164.23 / 5.70
& 76.87 / 58.67
& 3.45 / 3.09
& 387.92 / 131.28 \\

& Speculative Decoding
& 1.21 / \textbf{77.48}
& 5.07 / 2.45
& 117.58 / 21.46
& 501.14 / 5.49
& 625.00 / 106.88 \\

\bottomrule
\end{tabular*}

\vspace{1mm}
\begin{minipage}{\textwidth}
\footnotesize
\textit{Note:} In Panels A, B and D, entries are reported as TTFT / TPOT, Speculative Decoding method adopted is EAGLE-3. In Panel C, UL/DL denote Uplink/Downlink, Decode UL/DL shows Avg./Max. where applicable. Data sizes use 1~KB = 1024 bytes, 1~MB = 1024~KB, and 1~GB = 1024~MB. The overall accuracies on ARC-Easy, PIQA, and SciQ are 80.07\% for Qwen2-1.5B-Instruct and 89.73\% for Llama-3.1-8B-Instruct. Speculative Decoding uses 59 draft-tree nodes excluding the root. $\bar a = 2.39$, $\bar d = 3.97$, and $\alpha = 60.22\%$ (computed from unrounded means). 
\end{minipage}
\end{table*}

\subsection{Case Studies: A Metric-Driven Paradigm Comparison}
Table~\ref{tab:comprehensive_latency} presents a comprehensive stage-wise latency and transmission breakdown across the four paradigms. Rather than evaluating the paradigms in isolation, we analyze their viability through the lens of critical performance metrics.

\textit{1) Responsiveness and Network Sensitivity (TTFT).}
TTFT is highly sensitive to both network impairments and tensor transmission payloads. Under normal conditions, baseline Edge Inference achieves a TTFT of 73.13 ms. Under combined delay jitter and packet loss, end-to-end TTFT can reach approximately $4.4\times$ the normal-network baseline: the evaluated hidden-state-based Speculative Decoding implementation increases from 613.39 ms to 2692.26 ms under the same workload. Paradigms requiring heavy tensor transfers suffer substantial TTFT inflation. Model Split incurs a 367.10 ms TTFT, driven by a 175.19 ms uplink transmission of intermediate hidden states (2.3 MB). Under packet loss, this uplink phase increases to 345.60 ms. Similarly, the evaluated hidden-state-based implementation suffers a substantial downlink penalty during prefill: transmitting target-model hidden states (6.3 MB) consumes 491.37 ms, exceeding the edge compute time (110.93 ms).

We further consider a device--edge--cloud extension, where edge and cloud nodes serve requests with reusable KV caches. In this setting, jointly accounting for network latency, cache reuse, and serving load in request routing reduces the network-induced increase in mean TTFT by approximately 58\% relative to Least Load, which selects nodes solely according to their current load, in the studied scenario.

\textit{2) Generation Efficiency (TPOT).}
While Speculative Decoding theoretically accelerates autoregressive generation by reducing edge-side steps, our testbed reveals a counterintuitive wireless penalty. Under normal network conditions, its TPOT (63.61 ms) actually \textit{exceeds} the Edge Inference baseline (41.20 ms). The end-side draft generation (28.45 ms) combined with continuous transmission overhead negates the algorithmic speedup in the evaluated implementation. Model Split yields the worst TPOT (103.73 ms), with end-side and edge-side computation contributing 33.25 ms and 59.39 ms, respectively.
For Model Routing, requests assigned to Qwen2-1.5B-Instruct exhibit lower latency than those assigned to Llama-3.1-8B-Instruct in the evaluated workload. The smaller-model branch achieves a TTFT of 50.27 ms and a TPOT of 26.85 ms, compared with 74.66 ms and 40.95 ms for the larger-model branch, respectively.
The resource-constrained Jetson configuration further exposes sensitivity to end-side compute availability. For Model Split, end-side TTFT increases from 107.82 ms to 143.36 ms, and end-side TPOT rises from 33.25 ms to 63.83 ms. For the evaluated hidden-state-based implementation, end-side TPOT increases from 28.45 ms to 77.48 ms.

\textit{3) Communication Overhead and Bandwidth Constraints.}
Panel C of Table~\ref{tab:comprehensive_latency} highlights the stark contrast in data-plane requirements. Edge Inference and Model Routing are highly bandwidth-efficient, requiring only small prompt and token-streaming payloads. In contrast, the evaluated hidden-state-based Speculative Decoding implementation transfers 6.3 MB of target-model hidden states downlink during prefill, while Model Split transfers 2.3 MB of intermediate hidden states uplink per prefill. These massive tensor payloads risk saturating constrained 6G cellular links, confirming that without aggressive feature compression or dedicated high-bandwidth network slicing, these paradigms remain practically non-viable for resource-constrained terminals.
Directional rate limits further expose these communication bottlenecks. Under the 50 Mbps uplink limit, the Model Split uplink component of TTFT increases from 175.19 ms to 408.46 ms. Under the 50 Mbps downlink limit, the corresponding downlink component in the evaluated hidden-state-based implementation increases from 491.37 ms to 1618.77 ms.

\subsection{Context-Aware Tradeoffs of Edge-Device Collaboration}
\begin{figure}[!t]
\centering

\includegraphics[width=\columnwidth]{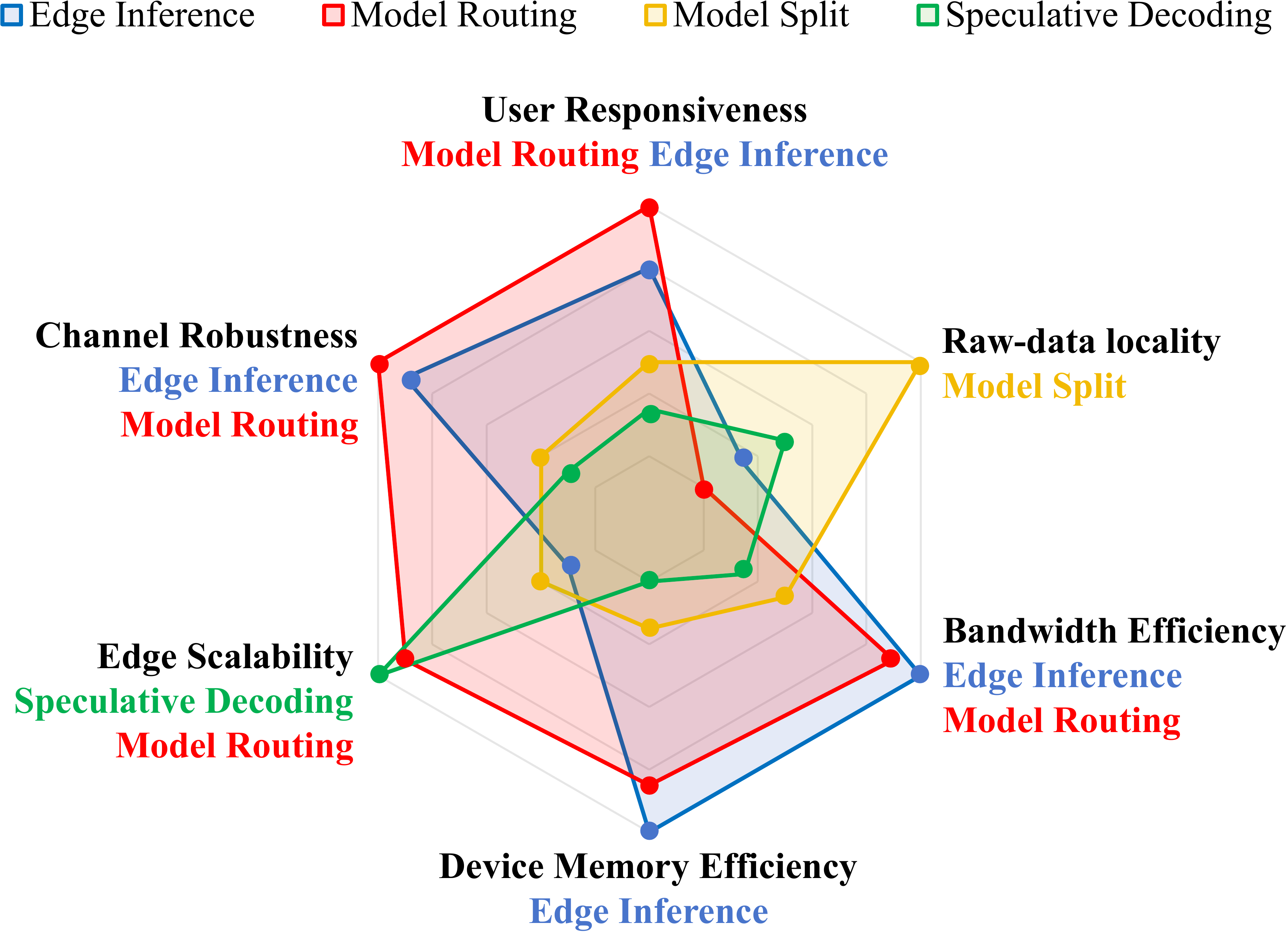}

\vspace{0.15cm}

\includegraphics[width=\columnwidth]{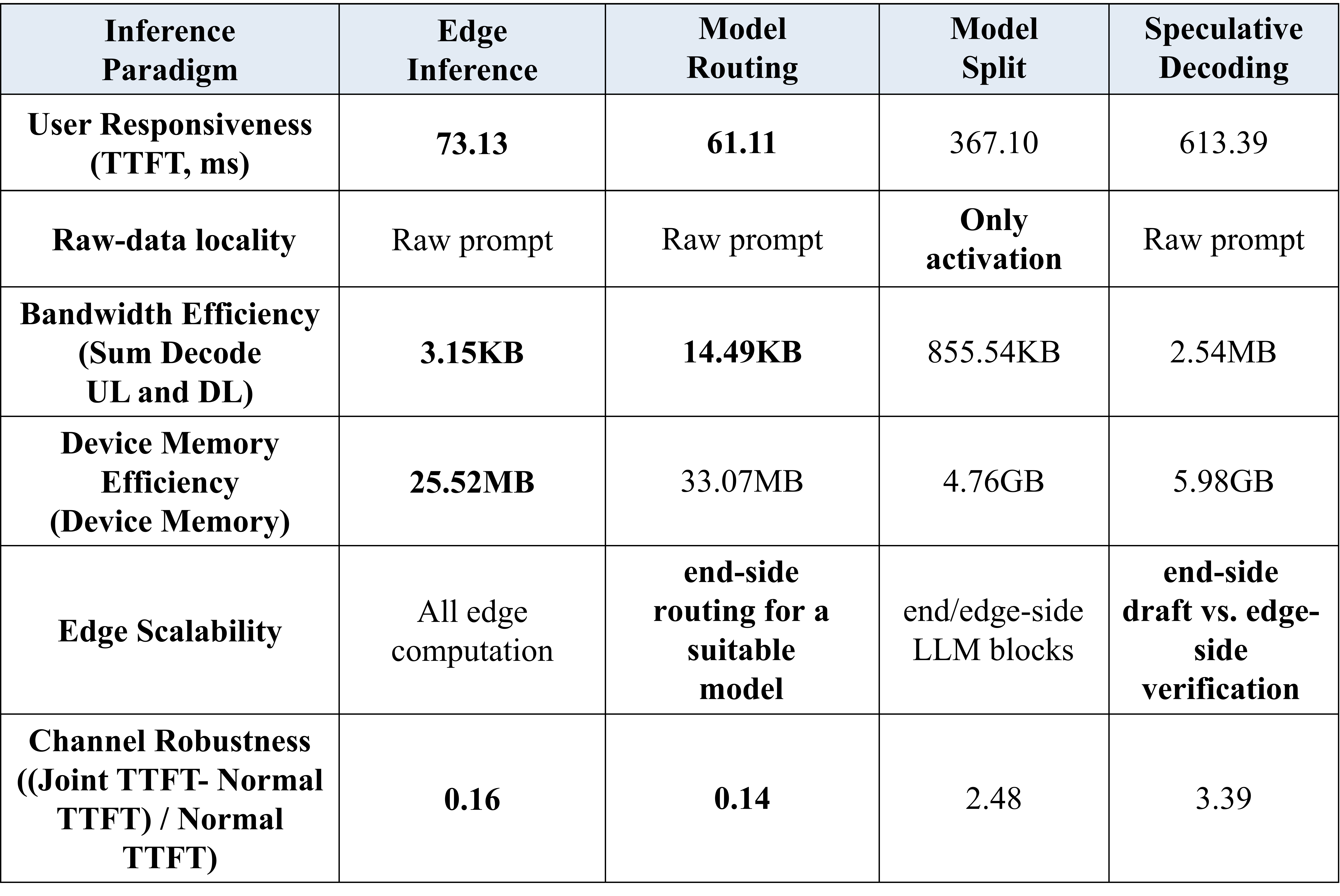}

\caption{Radar diagram and comparison matrix of the four collaboration paradigms. The analysis reveals that no single paradigm universally dominates the edge-device continuum.}
\label{fig:Radar}
\end{figure}

To holistically visualize the multidimensional trade-offs exposed by our testbed, we construct a radar comparison (Fig.~\ref{fig:Radar}) across six 6G-relevant axes: \textit{User Responsiveness} (TTFT), \textit{Raw-data locality}, \textit{Bandwidth Efficiency}, \textit{Device Memory Efficiency}, \textit{Edge Scalability}, and \textit{Channel Robustness}.
The synthesis conclusively demonstrates that no single paradigm universally dominates the 6G edge-device continuum:
\begin{itemize}
    \item \textbf{Edge Inference and Model Routing} excel in \textit{User Responsiveness} and \textit{Bandwidth Efficiency}, making them suitable choices for interactive, latency-sensitive applications (e.g., voice assistants). Model Routing can further improve edge resource utilization by distributing different types of requests among suitable models in the edge model pool.
    \item \textbf{Model Split} maximizes \textit{raw-data locality} by ensuring raw data never leaves the terminal domain. However, it scores low on \textit{User Responsiveness} and \textit{Channel Robustness} due to severe uplink tensor bottlenecks, making it relevant to applications with strict raw-data locality requirements where feature compression is available.
    \item \textbf{the hidden-state-based Speculative Decoding} offers superior \textit{Edge Scalability} by reducing edge-side autoregressive steps, but its \textit{Bandwidth Efficiency} and \textit{Channel Robustness} collapse under realistic wireless constraints due to massive downlink hidden-state transfers.
\end{itemize}
Notably, the ``optimal'' collaboration strategy is strictly context-dependent, dictated by a complex interplay of real-time network bandwidth, terminal compute capabilities, task semantics, and raw-data locality requirements. This empirical reality renders static paradigm selection fundamentally inadequate for 6G intelligent terminals, thereby necessitating the adaptive, context-aware orchestration framework.

\section{Context-aware Orchestration Framework and Future Directions}
The case studies conclusively demonstrate that static paradigm selection is inadequate for the heterogeneous realities of the 6G AI Edge. 
To bridge this gap, the network must transition toward intelligent, context-aware orchestration, paving the way for 6G edge-device collaboration.

\subsection{Context-aware Orchestration Framework}
To reconcile the conflicting demands of latency, raw-data locality, and bandwidth, we design a closed-loop, context-aware orchestration framework that dynamically selects or merges collaboration paradigms in real time. By shifting the complexity burden from the user to the network, this framework operates through a seamless three-phase workflow:

\textit{Phase 1: Task Requirement Identification (End-Side).}
Upon request initiation, the application identifies task-specific requirements, including \textit{Latency Tolerance} (e.g., tighter response-time requirements for interactive tasks than for background summarization), \textit{raw-data locality requirements} (e.g., keeping raw input on the device), and \textit{Task Type} (e.g., scientific question answering, everyday practical reasoning, programming, or economics). These requirements serve as inputs to the collaboration decision.

\textit{Phase 2: Context Aggregation and Decision Mapping (Edge-Side).}
The edge orchestrator combines task requirements with network and computing conditions, including RTT, jitter, uplink/downlink bandwidth availability, and edge computing load. A lightweight decision engine uses this information to select a suitable collaboration paradigm:
\begin{itemize}
    \item \textit{Scenario A (Strict raw-data locality + Sensitive Data):} The orchestrator selects \textbf{Model Split} to retain raw inputs on the terminal. Adaptive split-layer selection and feature compression (e.g., Fourier-based) could further mitigate uplink bottlenecks.
    \item \textit{Scenario B (Low Latency + Public Query):} For low-latency public queries without a model-matching requirement, the orchestrator selects \textbf{Edge Inference}, with latency-aware KV cache routing as a means to reduce edge queueing delay.
    \item \textit{Scenario C (Diverse Request Types + Multiple Edge Models):} The orchestrator employs \textbf{Model Routing} to match each request type to a suitable edge model. When multiple models are suitable for the same request type, serving load and latency can be considered as additional selection criteria. The \textbf{hidden-state-based Speculative Decoding} may be considered separately when the terminal has sufficient thermal and compute headroom for drafting.
\end{itemize}

We have conducted a small-scale offline evaluation of a requirement-aware decision tree using existing testbed measurements. Seven representative cases cover two Jetson configurations and six network profiles, with Model Split, Edge Inference, or Model Routing selected according to request requirements. Comparing the selected paradigms' measured TTFT with fixed Model Split shows a latency advantage in cases without a raw-data locality requirement, while retaining Model Split where locality is required. This comparison demonstrates the decision tree's advantage in balancing latency and request constraints through requirement-aware paradigm selection.

\textit{Phase 3: Seamless Execution and Mid-Session Fallback.} 
The edge provisions the selected pipeline and streams tokens to the user while the orchestrator continuously monitors the wireless channel. Should a sudden deep fade or mobility-induced handoff degrade bandwidth mid-stream, the framework dynamically triggers a fallback mechanism. For instance, when raw-input transfer is permitted, a bandwidth-heavy Model Split session could fall back to Edge Inference to mitigate generation stalls; otherwise, the fallback must preserve the raw-data locality requirement.

\subsection{Technical Roadmap for Edge-Device Collaboration}

\textit{1) Network-Aware KV Cache Orchestration:}
With TTFT reaching approximately $4.4\times$ the normal-network baseline under combined delay jitter and packet loss, \textit{predictive KV cache prefetching} based on mobility and request semantics could mitigate latency. \textit{Cross-edge KV cache federation} would enable MEC servers to reuse cached computations and route requests to nodes with pre-warmed contexts. Energy-aware scheduling should balance these latency benefits against terminal battery consumption.

\textit{2) Cross-Paradigm Tensor and Feature Compression:}
Fourier-based spectral compression~\cite{ma2025fouriercompress} and layer-aware quantization could reduce Model Split's uplink feature payloads. For the evaluated hidden-state-based Speculative Decoding implementation, compressing target-model hidden states could reduce downlink overhead. \textit{Adaptive compression ratios} should respond to channel conditions while balancing inference quality and transmission latency.

\textit{3) Stateful Session Management and Seamless Paradigm Migration:}
Mid-session migration requires \textit{stateful session management} to preserve model execution state, supported by interoperable model-state exchange and KV-cache serialization across heterogeneous edge platforms. Fallback from Model Split to Edge Inference must remain conditional on permission to transfer raw inputs; otherwise, migration must preserve raw-data locality.

\subsection{Future Directions of 6G AI Edge}
While current research predominantly focuses on collaborative LLM inference, the evolution of the 6G AI Edge will necessitate entirely novel collaboration paradigms:

\textit{1) Autonomous Agentic Inference.} 
Next-generation LLMs are evolving from passive text generators into autonomous agents capable of multi-step reasoning and tool utilization (e.g., ReAct frameworks). 
This necessitates a new \textit{Control-Plane Collaboration} paradigm. The edge server acts as the Brain, handling high-level planning and complex reasoning, while the terminal acts as the Body, executing local tool calls (e.g., reading local files, querying local sensors, or interacting with onboard APIs). 
Future research must design low-latency, bidirectional agentic loops that securely sandbox edge-generated code execution on the terminal.

\textit{2) Multi-Modal and Embodied AI Collaboration.} 
As 6G empowers autonomous robots and AR/VR headsets, terminals will continuously ingest massive multi-modal data streams (e.g., HD video, LiDAR point clouds). 
Transmitting raw multi-modal patches to the edge may saturate 6G uplinks. The future lies in \textit{Cross-Modal Split Inference}, where terminals extract and compress multi-modal embeddings locally, transmitting only the semantic representations to the edge. 
For embodied AI, edge-device collaboration must integrate with real-time learning, allowing the edge to simulate physics environments while the terminal executes low-latency motor-control policies.

\textit{3) Semantic Token Communication.} 
Recent studies have investigated Token Communication (TokCom) as an
emerging AI-native communication paradigm, including large-model-driven
cross-modal context-aware semantic communication, explainable
task-oriented token transmission, and token sharing among heterogeneous
language models~\cite{qiao2025tokcom, jiang2026explainable, fu2026tokcom}. Moving beyond treating wireless channels as dumb pipes for serialized bits (e.g., raw KV-cache tensors or token IDs), 6G will integrate LLM architectures directly with the physical layer.
The 6G vision of \textit{Semantic Token Communication} proposes bypassing traditional bit-level encoding entirely. Instead of protecting every bit of a hidden-state tensor with heavy channel coding, the terminal and edge can map LLM embeddings, draft tokens, or hidden features directly into semantic radio symbols. In this paradigm, channel noise does not cause catastrophic packet drops; rather, it introduces graceful semantic degradation, similar to how human communication handles noise, realizing the vision of a truly semantics-aware computing network.

\section{Conclusion}
Realizing the vision of 6G ubiquitous intelligent terminals requires edge-device collaborative LLM inference to bridge the gap between escalating model complexity and constrained mobile hardware. 
Through a unified, hardware-in-the-loop testbed, this article presents the first side-by-side empirical evaluation of four collaboration paradigms under realistic wireless conditions. Our stage-wise analysis shows that network impairments and tensor transmission overhead can increase TTFT to approximately 4.4 times its normal-profile value. The broader comparison reveals trade-offs among latency, communication overhead, and raw-data locality. 
This reality demands a definitive shift from static algorithmic deployment to dynamic, context-aware orchestration. 
Looking ahead, 6G AI Edge will require cross-layer innovations to empower a practical, scalable platform for next-generation AI applications.

\bibliographystyle{IEEEtran}
\bibliography{references}

\end{document}